\documentclass[final            
  ]
  {aipproc}

\layoutstyle{8x11double}

\begin{document}
\title{Leptogenesis in the E$_6$SSM: Flavour Dependent Lepton Asymmetries}


\author{S. F. King}{
  address={School of Physics and Astronomy, University of Southampton,
Southampton, SO17 1BJ, U.K.}
}

\author{\underline{R. Luo}}{
  address={Department of Physics and Astronomy, University of Glasgow,
Glasgow, G12 8QQ, U.K.}
}

\author{D. J. Miller}{
  address={Department of Physics and Astronomy, University of Glasgow,
Glasgow, G12 8QQ, U.K.}
}

\author{R. Nevzorov}{
address={Department of Physics and Astronomy, University of Glasgow,
Glasgow, G12 8QQ, U.K.}
}

\begin{abstract}
We discuss flavour dependent lepton asymmetries in the Exceptional Supersymmetric Standard Model (E$_6$SSM). In the E$_6$SSM, the right-handed neutrinos do not participate in gauge interactions, and they decay into leptons and leptoquarks. Their Majorana nature allows violation of lepton number. New particles and interactions can result in substantial lepton asymmetries, even for scales as low as $10^6\,{\rm GeV}$.
\end{abstract}

\maketitle


\section{Introduction}
The generation of the Baryon Asymmetry of the Universe (BAU) is a long-standing problem of particle cosmology. One plausible solution is thermal leptogenesis, where the Majorana right-handed (RH) neutrinos decay out-of-equilibrium into left-handed lepton doublets. CP violation in these decays leads to a non-zero lepton asymmetry as the universe cools down. {\it Sphaleron} processes convert part of this lepton number asymmetry into a baryon asymmetry in the universe before nucleosynthesis. 

However, in the seesaw scenario, the Yukawa couplings of RH neutrinos influence the light neutrino masses, constraining the generation of lepton asymmetry. In order to generate a sufficient lepton asymmetry in the Standard Model (SM) or the Minimal Supersymmetric Standard Model (MSSM), the mass of the lightest RH neutrino, $M_1$, should be $10^9\,{\rm GeV}$ or larger. On the other hand, the reheating temperature $T_R$ should be of order of $M_1$ because RH neutrinos are produced thermally. The reheating temperature is constrained by the production of gravitinos in supergravity models, which requires $T_R < 10^7\,{\rm GeV}$. This tension in the choice of the reheating temperature is known as the gravitino overproduction problem. Therefore it is interesting to consider leptogenesis in extended supersymmetric models; here we investigate the E$_6$SSM \cite{King:2008qb}. 

\section{E$_6$SSM}
The E$_6$SSM \cite{King:2005my}\cite{King:2005jy} is based on gauge group $SU(3)_c\times SU(2)_W\times U(1)_Y \times U(1)_N$ at low energies, which can originate from the breaking of an $E_6$ GUT theory. The extra $U(1)_N$ is a combination of $U(1)_\chi$ and $U(1)_\psi$ 
\begin{eqnarray}
U(1)_N=\frac{1}{4}U(1)_\chi +\frac{\sqrt{15}}{4} U(1)_\psi\,,
\end{eqnarray}
where $U(1)_\chi$ and $U(1)_\psi$ appear from the decomposition of $SO(10)$ and $E_6$ respectively. RH neutrinos are singlets with respect to $U(1)_N$ and therefore may gain large masses. The particle content of E$_6$SSM includes three fundamental representations of E$_6$. Each $27$-plet contains a generation of SM leptons and quarks, the RH neutrino $N_i$, two SU(2) doublets $H_{1,i}$ $H_{2,i}$ with quantum numbers of Higgs fields, a singlet field under the SM gauge groups $S_i$ and triplets of exotic quarks $D_i$ and $\overline D_i$, where $i=1,2,3$ is the generation index. Here, the third generation of $SU(2)$ doublets $H_d \equiv H_{1,3}$ and $H_u \equiv H_{2,3}$ play the role of the down-type and up-type Higgs fields respectively. The exotic quarks can be either diquarks, carrying 2/3 baryon number or leptoquarks carrying 1 lepton number and 1/3 baryon number (refered to as model I and model II respectively). The anomalies are canceled within each generation. In addition, extra doublets (only one generation) $L_4$ and $\overline L_4$ with mass $\sim 1\,{\rm TeV}$ are introduced in order to achieve gauge unification. $L_4$ effectively carries one lepton number and couples to SM leptons via Yukawa couplings. In order to suppress Flavour Changing Neutral Currents (FCNC) processes, a $Z_2^H$ symmetry is postulated, under which $H_{1,3}$, $H_{2,3}$, $S_3$ are even and all other fields are odd. The E$_6$superpotential invariant under a $Z_2^H$ symmetry is,
\begin{eqnarray}
W_{\rm E_6SSM}&=& \lambda S (H_u H_d)+ \lambda_{\alpha\beta} S (H_{1\alpha} H_{2\beta})+
\kappa_{ij} S (D_i\overline{D}_j)\nonumber\\ &&+ f_{\alpha\beta}(H_d H_{2\alpha})S_{\beta}
+\tilde{f}_{\alpha\beta}(H_{1\alpha}H_u)S_{\beta}\nonumber \\&&+ h^U_{ij}(H_{u} Q_i)u^c_{j} + h^D_{ij}(H_{d} Q_i)d^c_j\nonumber \\
&&+ h^E_{ij}(H_{d} L_i)e^c_{j}+ h_{ij}^N(H_{u} L_i)N_j^c\nonumber\\
&&+ \frac{1}{2}M_{ij}N^c_iN^c_j+\mu'(L_4 \overline{L}_4)\nonumber \\ && + h^{E}_{4j}(H_d L_4)e^c_j+h_{4j}^N (H_{u} L_4)N_j^c\,.
\end{eqnarray}

The $Z_2^H$ symmetry can only be approximate in order to allow exotic quarks to decay; couplings which break $Z_2^H$ should be suppressed. Via $Z_2^H$ symmetry breaking, the RH neutrinos couple to all leptons and $H_{2,i}$ of all generations. In model II, couplings of RH neutrinos to leptoquarks and down-type quarks (three generations) are also allowed. The corresponding contribution to the superpotential reads:
\begin{equation}
W_N= h^N_{kxj}(H^u_{k} L_x)N_j^c + g^{N}_{kij}D_{k}d^{c}_{i}N^{c}_{j}
\end{equation}
where $x=1,2,3,4$ and $k,i,j=1,2,3$, and $H^u_k \equiv H_{2,k}$ with $h^N_{3ij} \equiv h^N_{ij}$ and $h^N_{34j} \equiv h^N_{4j}$. The first term only is present in model I while both terms are present in model II.

\section{Lepton Asymmetries}

The first step in the study of leptogenesis is the calculation of the lepton asymmetries. Let's consider the situation where the $Z_2^H$ symmetry is conserved. In this scenario, the RH neutrinos couple to the first three generations of leptons and $L_4$ with $H_u$. The decay channels include
\begin{eqnarray}\label{NDecay0}
N_1\to L_x+H^u_k\,,\quad N_1\to \widetilde{L}_x+\widetilde{H}^u_k\,,\nonumber\\
\widetilde{N}_1\to \overline{L}_x+\overline{\widetilde{H}}^u_k\,,\quad\widetilde{N}_1\to \widetilde{L}_x+ H^u_k\,,
\end{eqnarray}
where $k=3$ only. The flavour dependent lepton asymmetries originate from the interference of tree-level decay amplitudes and one-loop corrections. Calculating the one-loop diagrams we find
\begin{equation}
\varepsilon^{3}_{1,\,\ell_x}\simeq-\frac{3}{8\pi}\sum_{j=2,3}
\frac{\mbox{Im}\biggl[(h^{N\dagger} h^{N})_{1j} h^{N*}_{x1} h^{N}_{xj}\biggr]}{(h^{N\dagger} h^{N})_{11}}\,\frac{M_1}{M_j}\,,
\end{equation}
where $M_{i=1,2,3}$ are the masses of RH neutrinos. We assume $M_1\ll M_{2,3}$ for all scenarios in this contribution.

When the effect of $Z_2^H$ symmetry breaking is considered, we should include the first and second generations of $H^u_k$ in the final states. The possible RH (s)neutrinos decay channels are given by Eq.(\ref{NDecay0}) with $k=1,2,3$, resulting in,
\begin{eqnarray}\label{lam1}
\varepsilon^{k}_{1,\,\ell_x}&=&\frac{1}{8\pi A_1}\sum_{j=2,3}\mbox{Im}\biggl\{
2\,A_j h^{N*}_{kx1} h^{N}_{kxj} \frac{M_1}{M_j}\nonumber\\
&&+\sum_{m,\,y} h^{N*}_{my1} h^{N}_{mxj}  h^{N}_{kyj} h^{N*}_{kx1}\,
\frac{M_1}{M_j}\biggr\},
\end{eqnarray}
where $A_j=\sum_{m,y} h^{N*}_{my1} h^{N}_{myj}$ and $x,y=1,2,3,4$, \,$i,k=1,2,3$.

It is convenient to define the {\it total} lepton asymmetries associated with each lepton flavour, 
\begin{eqnarray}
\varepsilon^{tot}_{1,\,f}=\sum_{k} \varepsilon^{k}_{1,\,f}\,.
\end{eqnarray}

In model II, leptoquarks may also contribute to the generation of lepton asymmetries. The additional RH neutrino decays are,
\begin{eqnarray}
N_1\to D_k+\widetilde{d^c}_i,\quad N_1\to \widetilde{D}_k+d^c_i\nonumber\\ \widetilde{N}_1\to \overline{D}_k+d_i,\quad 
\widetilde{N}_1\to \widetilde{D}_k+\widetilde{d^c}_i\,.
\end{eqnarray}
Notice that no baryon number is generated due to the presence of $D_k$ and $\tilde{d^c}_i$ in the final state, so only the lepton number violation need be considered. The lepton asymmetries with exotic quarks in the final states are,
\begin{eqnarray}
\varepsilon^{i}_{1,\,D_k}&=& \frac{1}{8\pi A_0}\sum_{j=2,3}\mbox{Im}\biggl\{
2\widetilde{A}_j g^{N}_{kij} g^{N*}_{ki1} \frac{M_1}{M_j} \nonumber\\
&&+ \sum_{m,\,n} g^{N*}_{mn1} g^{N}_{mij} g^{N}_{knj} g^{N*}_{ki1}
\frac{M_1}{M_j}\biggr\}\,,
\end{eqnarray}
where
\begin{eqnarray}
&\widetilde{A}_j=A_j+\frac{3}{2}\sum_{m,n} g^{N*}_{mn1} g^{N}_{mnj}\,,\\&A_0=\sum_{k,\,i}g^{N}_{ki1} g^{N*}_{ki1}\,.
\end{eqnarray} 
The new coupling of RH neutrinos and leptoquarks also contribute to $\varepsilon^{k}_{1,\,\ell_x}$. This can be described by Eq.(\ref{lam1}), but with $A_{2},\,A_3$ replaced by $\widetilde{A}_{2},\,\widetilde{A}_{3}$ separately. 

To illustrate how lepton asymmetries can be enhanced by these new couplings, we consider the generation of lepton decay asymmetries within see-saw models with sequential dominance \cite{King:2006hn}. The RH neutrino mass matrix and the Yukawa coupling matrix can be written as,
\begin{equation}
M_{N}=\left(
\begin{array}{ccc}
M_A & 0   & 0\\
0   & M_B & 0\\
0   & 0   & M_C
\end{array}
\right),\,
h^N=\left(
\begin{array}{ccc}
A_1 & B_1 & C_1\\
A_2 & B_2 & C_2\\
A_3 & B_3 & C_3
\end{array}
\right)\,.
\end{equation}
According to sequential dominance $\frac{|A_iA_j|}{M_A}\gg\frac{|B_iB_j|}{M_B}\gg\frac{|C_iC_j|}{M_C}$ and $|A_1|\ll|A_{2,3}|$. The left-handed neutrino masses can be obtained by diagonalizing the effective mass matrix in the basis of $(\nu_i,\,N_i)$
\begin{equation}
M=\left(
\begin{array}{cc}
0 & h^N\,v_u \\
h^{N\dagger}\,v_u   & M_N
\end{array}
\right),\,
\end{equation}
which gives,
\begin{equation}
m_3 \simeq \frac{(|A_2|^2+|A_3|^2)v_u^2}{M_A},\quad m_2\simeq \frac{|B_1|^2v_u^2}{s_{12}^2M_B}, \quad m_1\simeq 0 \,,
\end{equation}
where $v_u$ is the vev of Higgs field $H_u$ and $s_{12}\equiv\sin\theta_{12}$ is the sine of the 1-2 neutrino mixing angle. We further assume that $A_1=0$, $A_2=-A_3=|A|e^{i \phi_A}$ and $B_1=B_2=B_3=|B|e^{i \phi_B}$ as in Constrained Sequential Dominance. Note that $M_{A,B,C}\simeq M_{1,2,3}$, the RH neutrino masses, since diagonalizing the mass matrix gives a tiny contribution to the RH neutrino masses. Using the light neutrino masses $m_2\simeq8.7\times10^{-3}\,{\rm eV}$, $m_3\simeq 4.9\times10^{-2} {\rm eV}$ as in the case of a normal hierarchy, the relations outlined above provide all relevant Yukawa couplings, dependent only on the RH neutrino masses.

\begin{figure}[t]
    \includegraphics[width=.23\textwidth]{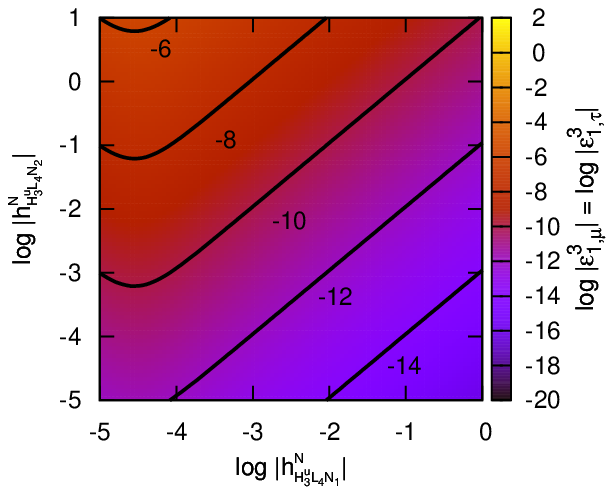}
  \includegraphics[width=.23\textwidth]{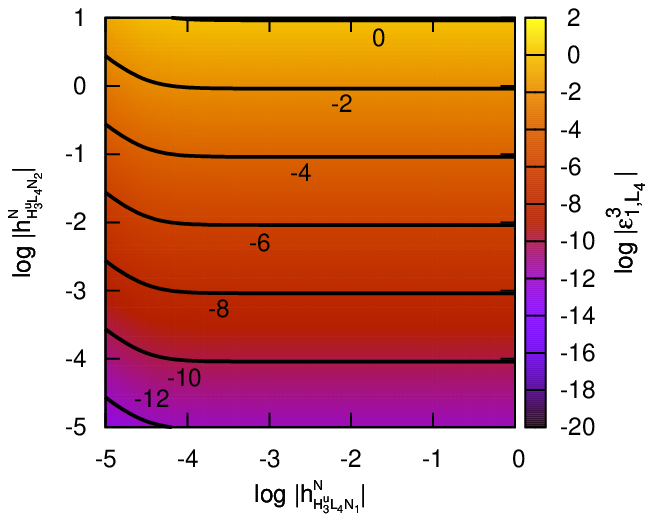}
  \caption{\label{fig1}Logarithm of the maximal values of $|\varepsilon^3_{1,\,\mu}|=|\varepsilon^3_{1,\,\tau}|$ (left) and $|\varepsilon^3_{1,\,L_4}|$ (right) with unbroken $Z_2^H$ symmetry versus $log|h^N_{H^u_3L_4N_1}|$ and $log|h^N_{H^u_3L_4N_2}|$ for $M_1=10^6\,{\rm GeV}$ and $M_2=10\cdot M_1$\,.}\label{figz2}
\end{figure}

\begin{figure}[t]
  \includegraphics[width=.23\textwidth]{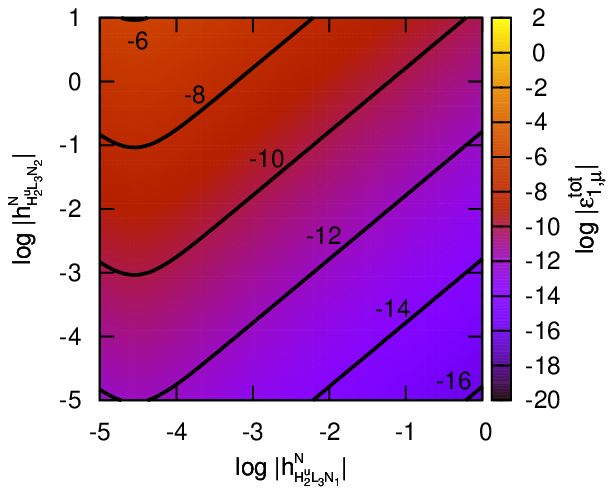}
    \includegraphics[width=.23\textwidth]{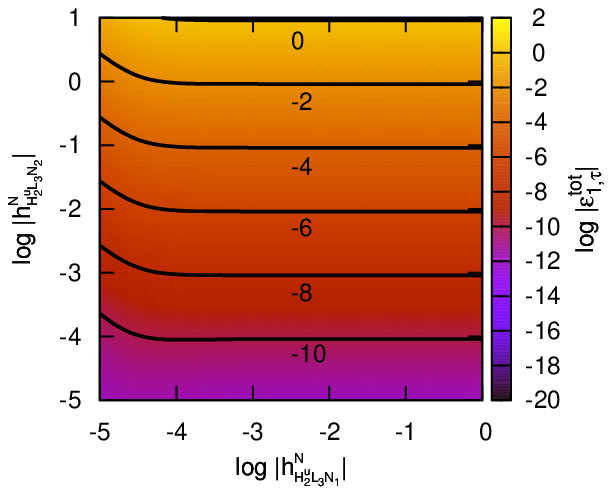}
  \caption{\label{fig1}Logarithm of the maximal values of $|\varepsilon^{tot}_{1,\,\mu}|$ (left) and $|\varepsilon^{tot}_{1,\,\tau}|$ (right) with broken $Z_2^H$ symmetry versus $log|h^{N}_{H^u_2 L_3 N_1}|$ and $log|h^{N}_{H^u_2 L_3 N_2}|$ for $M_1=10^6\,{\rm GeV}$ and $M_2=10\cdot M_1$\,.}\label{figm1}
\end{figure}

\begin{figure}[t]
  \includegraphics[width=.23\textwidth]{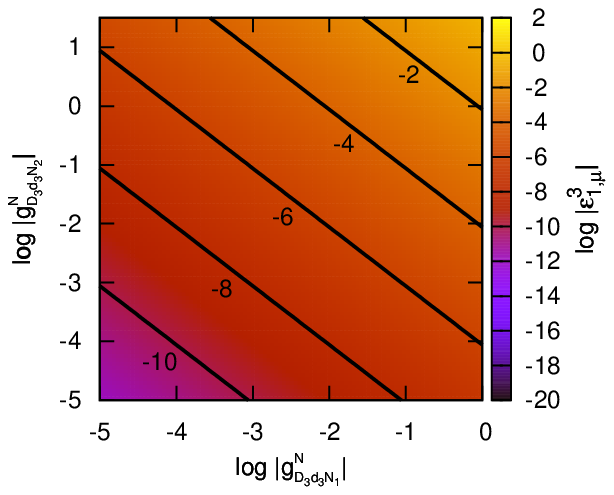}
    \includegraphics[width=.23\textwidth]{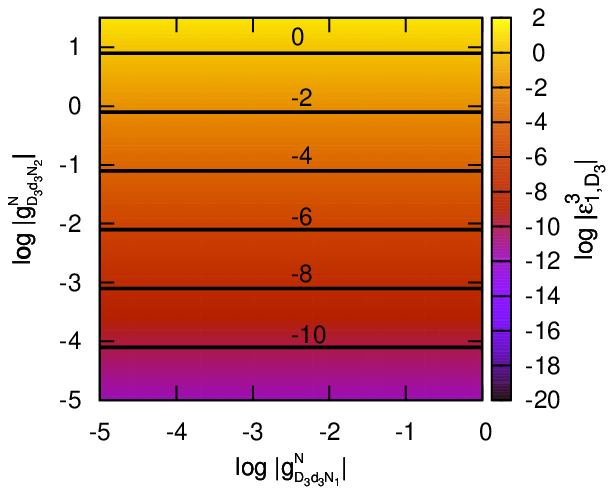}
  \caption{\label{fig1}Logarithm of the maximal values of $|\varepsilon^3_{1,\,\mu}|=|\varepsilon^3_{1,\,\tau}|$ (left) and $|\varepsilon^3_{1,\,D_3}|$ (right) and  with unbroken $Z_2^H$ symmetry versus $log|g^{N}_{D_3 d_3 N_1}|$ and $log|g^{N}_{D_3 d_3 N_2}|$ for $M_1=10^6\,{\rm GeV}$ and $M_2=10\cdot M_1$\,.}\label{figm2}
\end{figure}

We plot the maximal lepton decay asymmetries in Figs.(\ref{figz2}-\ref{figm2}). We adopt a notation where couplings of the form $h^N_{H^u_kL_xN_j}=h^N_{kxj}$ and $g^N_{D_3d_3N_i}=g_{33i}$. We estimate that in order to generate the observed baryon asymmetry with efficiency factor of $\eta\sim0.1$, we require lepton decays asymmetries of order $10^{-5}$.

In Fig.(\ref{figz2}) we restrict our consideration to the $Z^H_2$ symmetric case and ignore the couplings of the heaviest RH neutrino. One can see that a substantial CP asymmetry associated with $L_4$ can be achieved even for $M_1 = 10^6 \, {\rm GeV}$ if $h^N_{H^u_3 L_4 N_2}$ is of the order $10^{-1}$ -- $10^{-2}$. 

Figs.(\ref{figm1}) and (\ref{figm2}) illustrate scenarios with broken $Z^H_2$ symmetry. In this case, for simplicity, we ignore the Yukawa couplings of $L_4$ to the RH neutrino. In Fig.(\ref{figm1}) we examine the dependence of the maximal CP asymmetry in Model I, assuming that only $h^N_{H^u_2L_3N_1}$ and $h^N_{H^u_2L_3N_2}$ have non-zero values. Once again we see a substantial CP asymmetry can be generated if $h^N_{H^u_2 L_3 N_2}$ is of the order $10^{-1}$ -- $10^{-2}$. 

Finally in Fig.(\ref{figm2}) we consider the generation of CP asymmetries in model II. For simplicity, we ignore the couplings of all exotic particles except those of the third generation of leptoquarks $D_3$, which predominantly couples to the RH $b$-quark $d_3$ and the two lightest RH neutrinos, $N_1$ and $N_2$. As in the previous two cases, an appropriate amount of CP asymmetry can be generated by the couplings of the exotic quarks.

\section{Conclusions}
We calculate the flavour dependent lepton asymmetries in the E$_6$SSM. New particles and new sources of CP violation may result in the drastic enhancement of lepton decay asymmetries. We demostrate how the new couplings increase lepton CP asymmetries within seesaw models with Sequential Dominance. Successful leptogenesis can be achieved for a wide range of parameters. To provide a quantification of the resulting baryon asymmetry requires detailed study of the complete system of flavour dependent Boltzmann equations.

\bibliographystyle{aipproc}   

\bibliography{e6ssmlptg}

\IfFileExists{\jobname.bbl}{}
 {\typeout{}
  \typeout{******************************************}
  \typeout{** Please run "bibtex \jobname" to optain}
  \typeout{** the bibliography and then re-run LaTeX}
  \typeout{** twice to fix the references!}
  \typeout{******************************************}
  \typeout{}
 }

\end{document}